\documentclass[prl,twocolumn,superscriptaddress,showpacs,floatfix]{revtex4}
\usepackage{epsfig}

\begin{document}

\title{Surface Nucleation in the Freezing of Gold Nanoparticles. }

\author{Eduardo Mendez-Villuendas}
\affiliation{Department of Chemistry, University of Saskatchewan,
Saskatoon, Saskatchewan, S7N 5C9}

\author{Richard K. Bowles\footnote{Corresponding author email: richard.bowles@usask.ca}}
\affiliation{Department of Chemistry, University of Saskatchewan,
Saskatoon, Saskatchewan, S7N 5C9}
\email{richard.bowles@usask.ca}

\date{\today}

\begin{abstract}
We use molecular simulation to calculate the nucleation free energy barrier for the freezing of a 456 atom gold cluster over a range of temperatures. The results show that the embryo of the solid cluster grows at the vapor-surface interface for all temperatures studied and that the usual classical nucleation model, with the embryo growing in the core of the cluster, is unable to predict the shape of the free energy barrier. We use a simple partial wetting model that treats the crystal as a lens-shaped nucleus at the liquid-vapor interface and find that the line tension plays an important role in the freezing of gold nanoparticles.
\end{abstract}



\maketitle

Nanoclusters, consisting of only tens to thousands of molecules, exhibit a rich variety of structures and phase transitions that are very different from their bulk counterparts. The most energetically stable structure of a cluster varies as a function of the number of atoms or molecules. For small clusters, the Mackay or anti-Mackay icosahedra are usually the most stable structures, with the magic numbers corresponding to completed icosahedral shells being particularly stable. As the clusters become larger, decahedra and eventually the  face-centred-cubic (fcc) structures become energetically favorable~\cite{wales_book}.  However, molecular dynamics (MD) simulations of cooled clusters show that nanoclusters generally freeze to a metastable state rather than the energetically most favored one, suggesting kinetic factors play an important role in determining nanoparticle structure~\cite{baletto05}. For example, the fcc is the most stable structure for gold particles with more than 500 atoms but Bartell et al show that clusters with more than $N=1000$ still predominantly freeze to  icosahedra~\cite{bartell01}. 


Surface phenomena are expected to play an important role in the freezing of nanoparticles. Molecular dynamics simulations used to study the freezing of a 561 atom gold cluster, cooled at a constant rate from above the melting temperature, show the formation of the icoshahedral structure is initiated by ordering at the surface rather than in the core~\cite{nam02}. On the other hand, Lennard-Jones clusters initially freeze to a core-ordered icosahedron with a disordered surface~\cite{man06}. This suggests that the actual location (i.e. surface or core) of nucleation will be driven by the wetting behavior of the crystal and liquid interface. Understanding where nucleation takes place within the nanoparticle is important because homogeneous and heterogeneous (surface) nucleation rates can differ by orders of magnitude.

Furthermore, the location of the forming embryo influences the nature of the phenomenological models commonly used to determine the surface free energy densities of crystals. Recent simulations of crystal nucleation have shown that classical nucleation theory (CNT) and the usual capillarity-based models are not always sufficient to describe the free energy barrier~\cite{mixed_hs,wall_hs}. In the case of gold, it is the liquid-solid surface free energy density that is of interest and its value is generally obtained from models that assume the solid embryo forms in the core of the nanoparticle so that it is completely wet by the liquid~\cite{bartell01b}.

In this letter, we investigate the freezing of gold nanoparticles by calculating the free energy barrier to nucleation using Monte Carlo (MC) simulation techniques. Previously, Nam et al~\cite{nam_prb} calculated the free energy of a gold cluster with respect to a global order parameter $Q_{6}$, which characterises the total degree of crystallinity of the cluster, for a range of temperatures and found that the barrier between the liquid phase and the icosahedron was considerably lower than the corresponding barrier to the more stable fcc crystal. We have chosen to characterise the nucleation barrier in terms of the size of a solid-like embryo $n$ in order to gain insight into the molecular details of the nucleation process. This also allows us to compare our computed barrier heights with the results of CNT, which focuses on the thermodynamic work of forming an $n$-sized embryo. 

Our criterion for identifying an $n$-sized solid-like embryo in a cluster is a slightly modified version of the criterion used in studies of crystal nucleation in bulk systems~\cite{fre_order}. Following Frenkel, we begin by defining a 13-dimensional complex vector with components,
\begin{equation}
\label{bvec}
q_{6m}(i)=\frac{1}{n_{b}(i)}\sum_{j=1}^{n_{b}(i)}Y_{6m}(\hat{r}_{ij})\mbox{ ,}\\
\end{equation}
where the sum is over all the neighboring atoms, $n_{b}(i)$, within a radius of $3.5\AA$ of atom $i$. This distance usually contains about 12-13 neighbors if the atom is in the core of the cluster. $Y_{6m}(\theta,\phi)$ is the $6^{th}$ order spherical harmonic and $\hat{r}_{ij}$ is the unit vector pointing from particle $i$ to a neighbor $j$ that specifies the elevation and azimuth angles that this bond makes with respect to the coordinate system. Eq.~\ref{bvec} characterises the local order surrounding an individual particle but we also expect the local order of neighboring atoms in a solid embryo to be highly correlated so we consider two neighbors to be {\it connected} if the correlation function, $c_{ij}={\bf q}_{6}(i)\cdot{\bf  q}_{6}(j)=\sum_{m=-6}^{6}q_{6m}(i)\cdot q_{6m}^{*}(i)$,
where $^{*}$ denotes the complex conjugate, is above the threshold value of 0.65. This threshold value was obtained by comparing the distribution functions of $c_{ij}$ obtained for temperatures above and below the freezing temperature for the cluster $T=750$K\cite{melt} and selecting the point at which they intersect. The two distributions were not totally separated and while the intersect varied slightly, depending on the two selected temperatures, this generally occurred in the range $c_{ij}=0.6-0.7$. For a much larger cluster size ($N=3892$) the threshold value can be determined more accurately, and is 0.68\cite{melt}. To further distinguish between liquid- and solid-like particles we require at least half of the neighbors of a solid particle to be connected. This last point is where our criterion differs from the bulk model and the adjustment is required because there are many different environments within a cluster, and atoms in the core, surface or at a vertex all have very different numbers of neighbors. Finally, two solid atoms are considered to be in the same embryo if they are {\it connected.} Similar order parameters based on Eq.~\ref{bvec} have been used previously to identify solid-like particles in gold nanoclusters~\cite{bartell01b}.

 We study a $N=465$ atom gold cluster, using the semi-empirical embedded-atom method (EAM) potential~\cite{gold_p},  in the $N,V,T$ ensemble, with a cell volume $V=1500\AA^{3}$ and periodic boundary conditions, to be consistent with pervious nucleation studies~\cite{bartell01b,nam02}. To calculate the free energy barrier as a function of embryo size $n$, we implement an umbrella sampling scheme to ensure accurate sampling for each embryo size, coupled with parallel tempering in temperature~\cite{gey95,fre_order}.  At each temperature, we run eight parallel simulations or windows, each with a parabolic biasing potential $w(n_{max})=0.0005(n_{max}-n_{0})^2$ which biases the system to sample states where the largest embryo $n_{max}$ in the cluster is around $n_{0}$.  We choose $n_{0}=0,10,20,30\ldots70$  and use $T=750,730,710,690,680,670$ for tempering. The embryo criterion is computationally expensive to apply so we use trajectories that consist of 10 normal MC moves for every particle in the cluster sampling the EAM potential, followed by a test against $w(n_{max})$. If the final move is rejected, the system is returned to the state at the beginning of the trajectory.  During the simulation, we attempt to swap clusters with those in neighboring windows every 10 trajectories and accept swaps according to the usual replica exchange probabilities~\cite{gey95}. We also attempt switches in neighboring temperatures ($n_{0}$ fixed) every 10 trajectories, but these are offset with the $n_{o}$ switches. The tempering switches have acceptance ratios of about 0.4 and 0.6 respectively.


The work of forming an $n$-sized embryo, $W(n)$, is related to the probability, $P_{n}$, of observing the fluctuation by a Boltzmann weighted equation. If the system is only mildly supercooled, so that the appearance of an embryo is rare, then the equilibrium number of clusters, $N_{n}$, is approximately equal to $P_{n}$ so that~\cite{fre_order,bow99}
\begin{equation}
\label{work}
\frac{N_{n}}{N}\approx P_{n}=\exp[-W(n)/kT]\mbox{ ,}\\
\end{equation}
where $k$ is the Boltzmann constant. We measure the ensemble average of $N_{n}$  and use Eq. \ref{work} to obtain $W(n)/kT$ from each window. The free energies in each window differ by an unknown additive constant, so the full free energy curve is constructed fitting the curves to a polynomial in $n$~\cite{fre_order}.

\begin{figure}[b]
\includegraphics[width=3.5in]{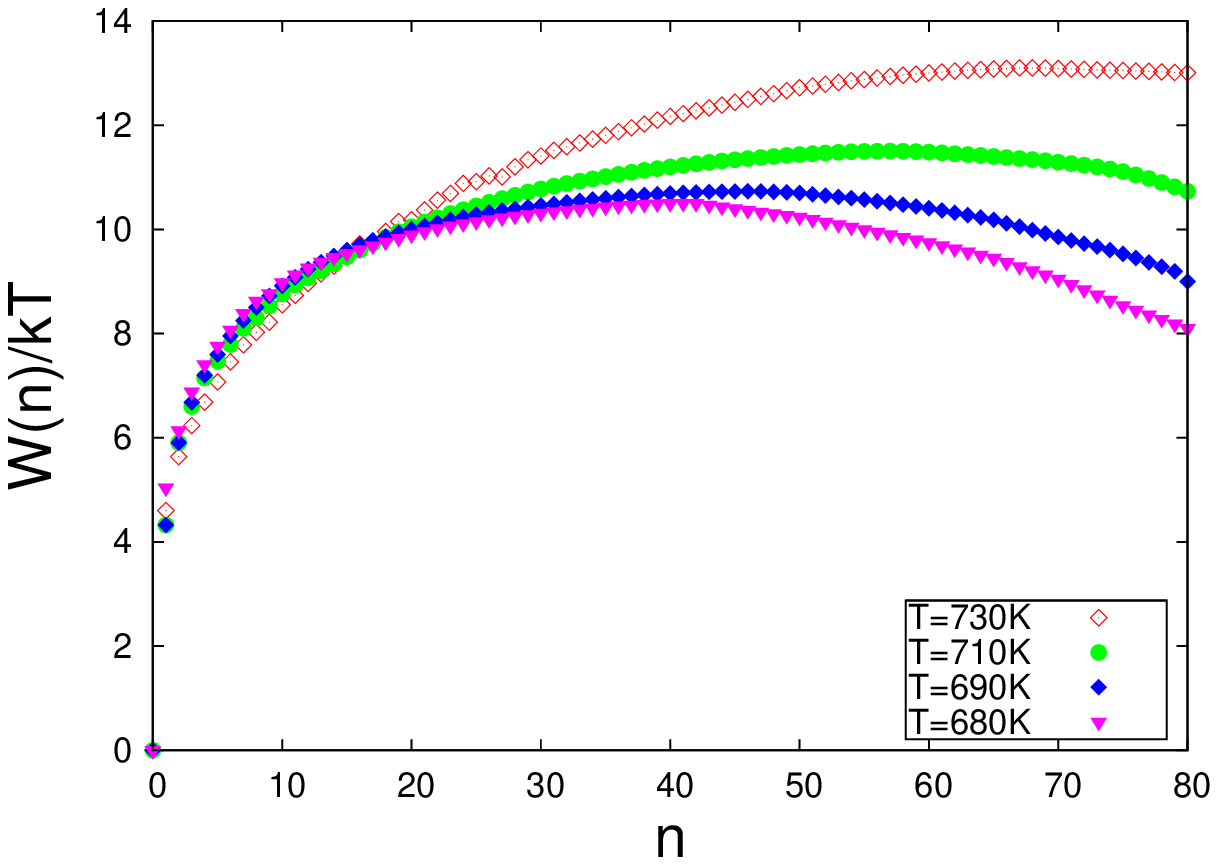}
\caption{Free energy of formation $W(n)/kT$ of an $n$-sized embryo for temperatures $T=730K\mbox{, } 710K, 690K\mbox{ and }680K$.}
\label{fig:barriers}
\end{figure}

Fig.~\ref{fig:barriers}  shows the free energy barriers for nucleation at four temperatures just below the melting temperature. These results were obtained as an average of four independent simulations with each sampling 436,000 trajectories after the system reached equilibrium i.e.~a total of $1.744\times 10^6$ trajectories. An estimate of the error was obtained by dividing the four runs in half to construct eight individual free energy curves and calculating the standard deviation of these curves from the average. This gives us an error of about $0.6kT$. 
The free energy curve shows the expected maximum and this decreases, along with the size of the critical nucleus, with increased supercooling. However, we note that for very small embryo sizes, $W(n)/kT$ is actually increasing with decreasing $T$. This feature would not be predicted using a simple CNT model.

\begin{figure}[h]
\label{Nsurf}
\includegraphics[width=3.2in]{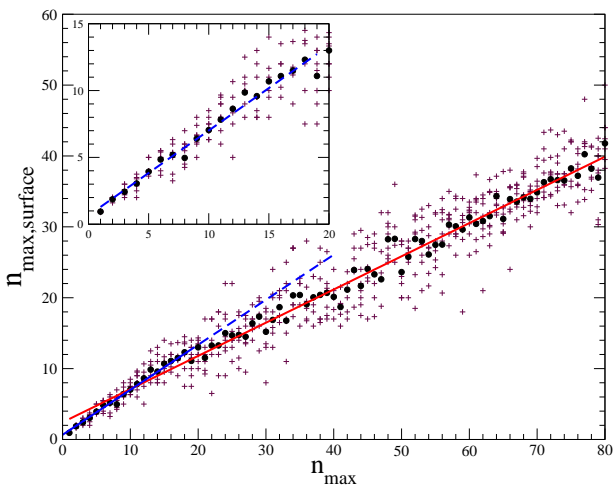}
\caption{The average number of surface atoms $n_{max,surf}$ found in the largest embryo $n_{max}$. The dark circles represent the values averaged over all temperatures. The lines are linear best fits for the larger (red) and smaller (blue) embryos. The cross symbols represent the averages for each temperature studied and give an indication of the scatter of the data. The insert is a magnification of the small embryo region. In color online.}
\label{fig:Nsurf}
\end{figure}

\begin{figure}[h]
\label{Nsurf}
\includegraphics[height=1.2in]{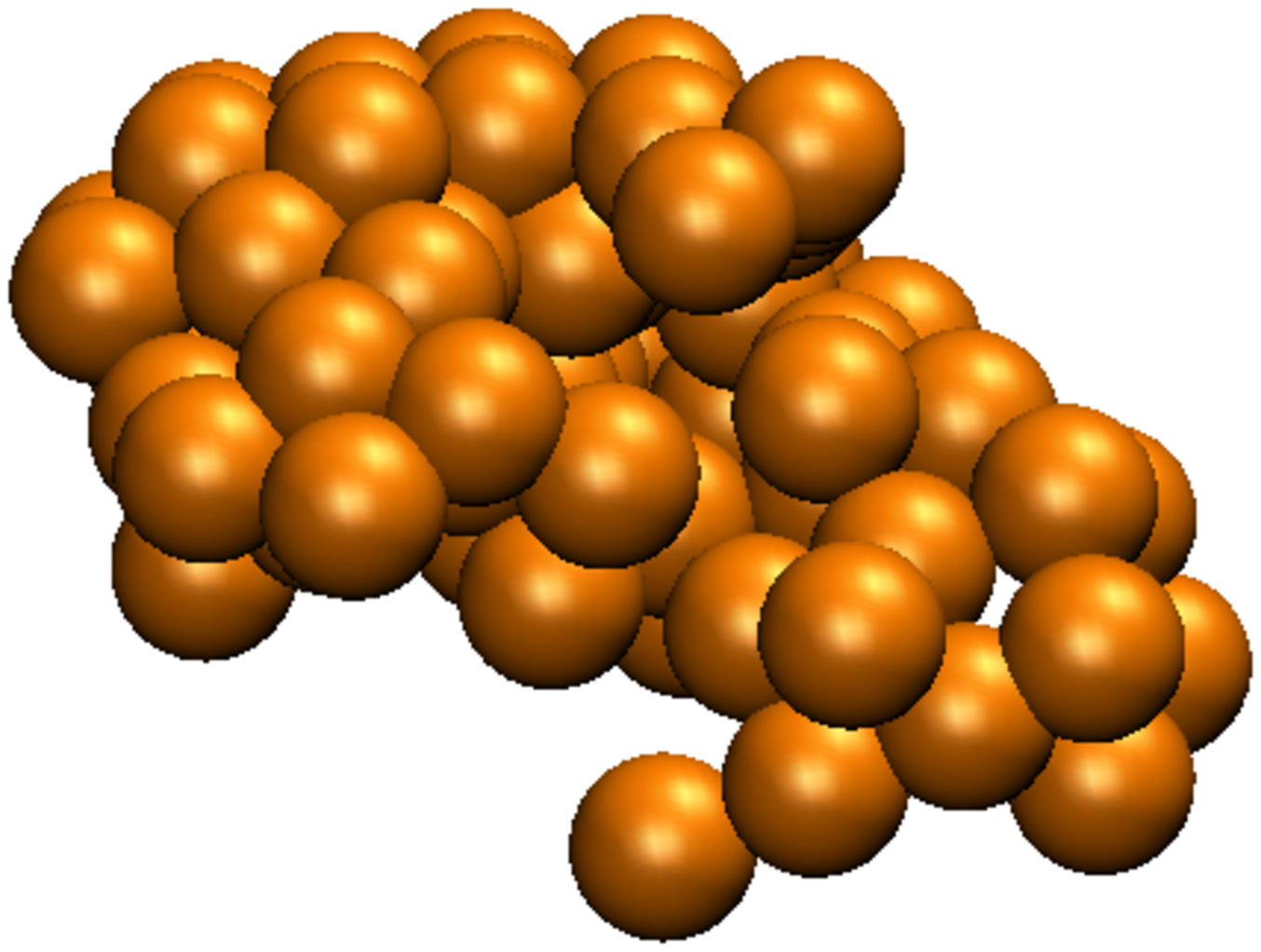}
\includegraphics[width=1.5in]{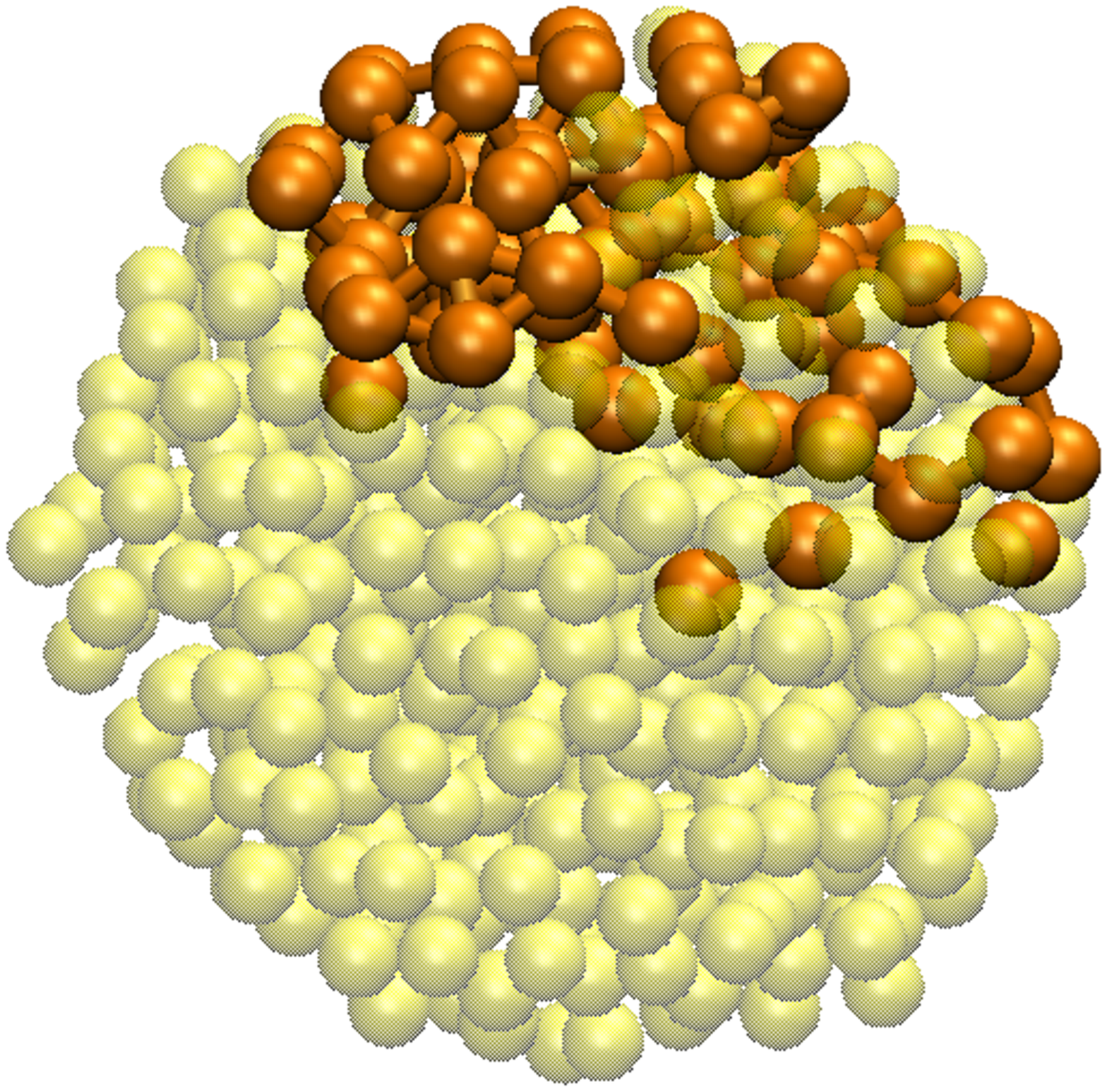}
\caption{Left: An isolated 71 atom embryo. Right: The same solid embryo (dark atoms) embedded in the liquid (light atoms) cluster. In color online. }
\label{fig:embryo}
\end{figure}

We are also interested in understanding where the embryo forms. In particular, we want to know whether it forms at the liquid-vapor interface or within the core of the cluster.  To identify surface atoms, we use a ``cone" algorithm~\cite{cone} with an apex angle of $120^{\circ}$. For a cluster of 456 atoms, approximately 52\% of the atoms are at the surface. Fig.~\ref{fig:Nsurf} shows the average number of atoms in the largest embryo found on the cluster surface, $n_{max,surface}$ versus the embryo size $n_{max}$. The cross symbols represent the values for all the temperatures studied and we have made these all the same color and symbol for clarity. The results for each of the individual temperatures can be fitted to a linear curve but all the lines are essentially the same within the scatter of the data so we have averaged over all temperatures to obtain the dark symbols. Approximately 46\% of the atoms in the larger embryos and 63\% of the atoms in the smaller embryos lie at the cluster-vapor interface, clearly suggesting that the growing solid phase is partially wet by the liquid. Fig.~\ref{fig:embryo} shows a typical large embryo.

The condition of partial wetting places thermodynamic limits on the surface free energy densities ($\sigma_{ij}$) of the three phases such that $\sigma_{13}-\sigma_{23}<\sigma_{12}$, where the subscripts 1, 2 and 3 denote the solid, liquid and vapor phases respectively~\cite{sn}. For the EAM potential, $\sigma_{13}=0.90J/m^{2}$ and $\sigma_{23}=0.74J/m^{2}$~\cite{st_eam} which requires $\sigma_{12}>0.16J/m^{2}$. Bartell et al~\cite{bartell01b} found that a number of thermodynamic theories and empirical relations gave estimates of $\sigma_{12}$ in the range of $0.11-0.16J/m^{2}$. In the same work, the authors used a CNT model that assumes complete wetting of the solid embryo by the liquid (core nucleation) to predict the solid-liquid surface tension based on fitting the rate of nucleation obtained from molecular dynamics simulations. For a cluster of $N=459$, at $T=720$K, they found $\sigma_{12}=0.084J/m^2$ which is well below the wetting threshold. 

\begin{figure}[t]
\includegraphics[width=3.2in]{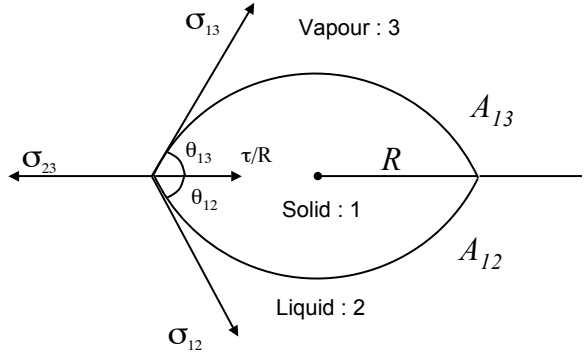}
\caption{Schematic cross section of a solid (phase 1) lens nucleus forming at the liquid (phase 2) - vapor (phase 3) interface. $A_{12}$ and $A_{13}$ are the solid-liquid and solid-vapor interfacial surface areas respectively and $R$ is the radius of the lens. The four arrows originating from the 3-phase contact are the force vectors of the surface tensions $\sigma_{ij}$ and the line tension $\tau/R$.}
\label{fig:lens}
\end{figure}

To obtain an estimate of $\sigma_{12}$ under the conditions of partial wetting, we assume the solid embryo grows at a planar liquid-vapor interface in the shape of a lens (see Fig.~\ref{fig:lens}). The lens model, without a line tension contribution, has previously been used to study droplet formation at the liquid-vapor interface~\cite{lens} and its application to crystal nucleation in a small nanoparticle implies a significant simplification of the microscopic process. We continue to ascribe bulk-like properties, such as chemical potential and surface tensions, to a nucleus containing just tens of atoms, in line with CNT, and we ignore many effects including surface roughness and the appearance of crystal facets. Nevertheless, such a model is numerically tractable and provides some useful insight into the nucleation process. Fits to our simulation data using the core nucleation CNT model, or the lens model without the line tension, were unable to account for the shape of the free energy curve.
Auer and Frenkel~\cite{wall_hs} found that the line tension, $\tau$, and its curvature correction, $\tau_0$, played an important role in the heterogeneous freezing of hard sphere colloids at a wall. Including $\tau$ in the mechanical equilibrium at the three phase contact line for the lens model makes the contact angles size-dependent~\cite{cline} so 
\begin{equation}
\label{c12}
\cos\theta_{12}=\frac{\sigma_{23}^{2}+\sigma_{12}^{2}-\sigma_{13}^{2}-2\sigma_{23}\tau/R+(\tau/R)^{2}}{2\sigma_{12}(\sigma_{23}-\tau/R)}\\
\end{equation}
and
\begin{equation}
\label{c13}
\cos\theta_{13}=\frac{\sigma_{23}^{2}+\sigma_{13}^{2}-\sigma_{12}^{2}-2\sigma_{23}\tau/R+(\tau/R)^{2}}{2\sigma_{13}(\sigma_{23}-\tau/R)}\mbox{ .}\\
\end{equation}

The work needed to form an $n$-sized embryo can then be expressed
\begin{equation}
\label{work2}
W(n)=n\Delta\mu+R^{2}\left[\sum_{i=2,3}\frac{2\pi\sigma_{1i}}{1+\cos\theta_{1i}}-\pi\sigma_{23}\right]+2\pi R\left[\tau+\frac{\tau_{0}}{R}\right]\\
\end{equation}
where $\Delta\mu$ is the difference in chemical potential between the liquid and solid phase. $R$ is obtained numerically under the constraint of fixed volume for an $n$-sized embryo,
\begin{equation}
nv=\frac{\pi}{3}R^{3}A_{0}\mbox{ : }A_{0}=\sum_{i=2,3}\frac{\sin\theta_{1i}(2+\cos\theta_{1i})}{(1+\cos\theta_{1i})^{2}}\\
\end{equation}
where $v=1.7277\times 10^{-29}m^{3}$ is the volume per molecule in the solid phase.

\begin{figure}[t]
\includegraphics[width=3.1in]{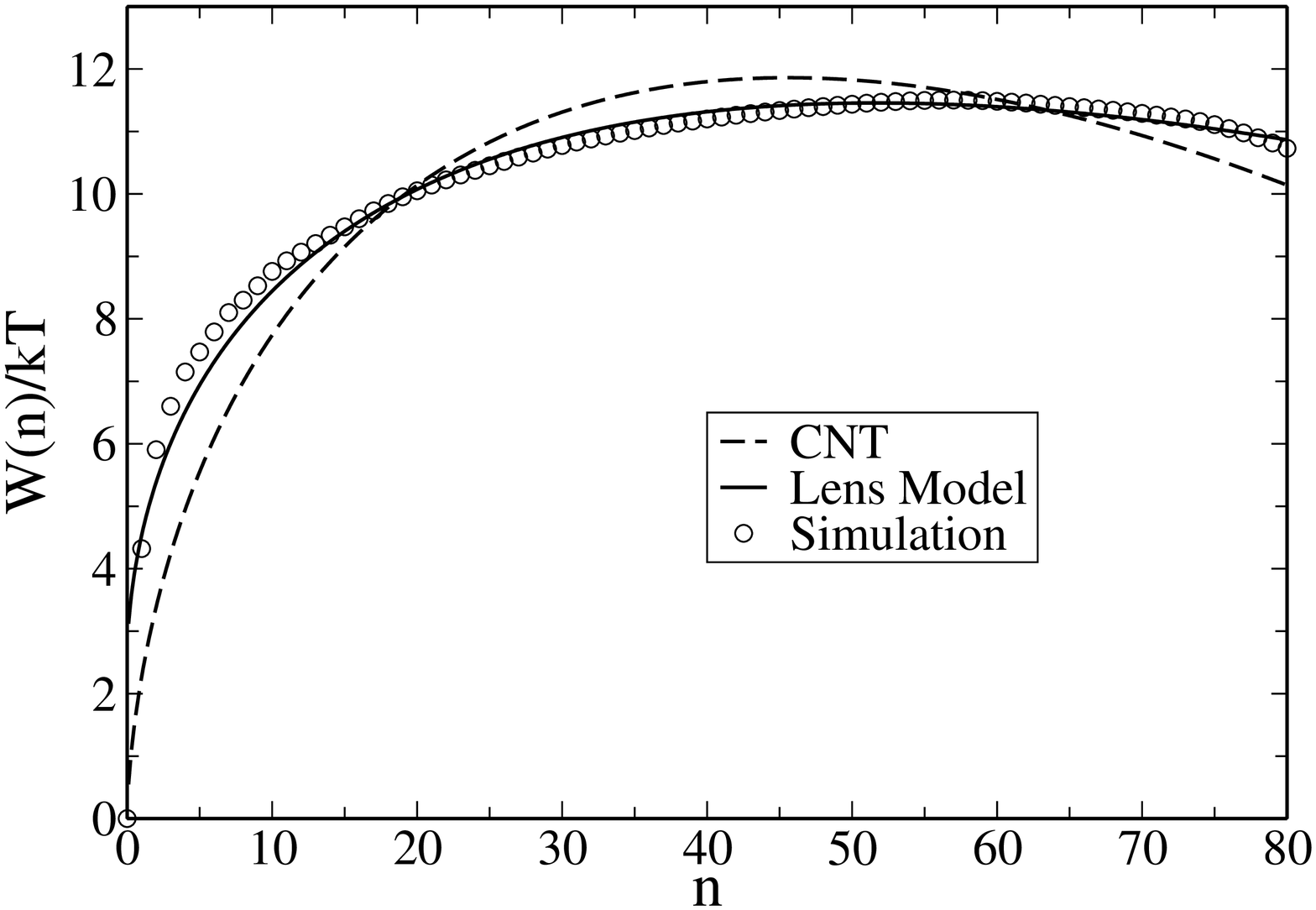}
\caption{Phenomenological model data fits to the calculated free energy barrier at $T=710K$. The lens model with line tension included (solid line) and CNT, assuming core nucleation (dotted line).}
\label{fig:fit_tau}
\end{figure}

Fig.~\ref{fig:fit_tau} shows a fit of the model to the data at $T=710$ where $\Delta\mu/kT=-0.22$, $\sigma_{23}=0.18J/m^{2}$, $\tau=-1.17\times 10^{-11} J/m$ and $\tau_{0}=3.92\times 10^{-21}J$ were adjustable parameters. Due to the nature of the numerical fit, we cannot guarantee that we are at the global minimum for the data fit and there is still the possibility of finding an improved fit with a positive line tension. At small embryo sizes, we can expect a greater fraction of particles to be in the surface of the lens based on surface-to-volume ratio arguments, and hence, a greater fraction of solid-like particles in the small embryos will appear on the surface of the cluster, as compared to the larger embryos.  A negative $\tau$ would enhance this effect by stretching the lens of small embryos, where line tension is most important. We also fit the CNT core nucleation model to our data at $T=710$K, using $\mu$ and $\sigma_{12}$ as adjustable parameters and assuming a spherical geometry for the embryo. The resulting $\sigma_{12}=0.085J/m^{2}$ is the same as that obtained from direct measurements of the rate~\cite{bartell01b}. We see in Fig.~\ref{fig:fit_tau} that the model clearly fails to predict the correct shape of the barrier but does obtain a close estimate of the barrier height.

\acknowledgements
We would like to thank I. Saika-Voivod for a number of useful discussions. We acknowledge NSERC for funding and WESTGRID for computing resources.




\end{document}